\documentclass[12pt]{IEEEtran}
\usepackage{graphicx, amssymb}
\usepackage{graphicx}
\newtheorem{thm}{Theorem}[section]
\newtheorem{lem}[thm]{Lemma}

\newtheorem{assum}[thm]{Assumption}

\makeatletter
\def\ifundefined{\@ifundefined}
\makeatother

\begin{document}

\title{Reduced Complexity Joint Iterative Equalization and Multiuser Detection in Dispersive DS-CDMA Channels}

\author{Husheng Li and H. Vincent Poor\thanks{Department of Electrical Engineering, Princeton University, Princeton, NJ 08544, USA ~(email:
\{hushengl, poor\}@princeton.edu). This research was supported in
part by the Army Research Laboratory under Contract DAAD
19-01-2-0011 and in part by the New Jersey Center for Wireless
Telecommunications.}}

% The major version number of the class file will not
% be defined with the old IEEEtran.cls. So, we can use this fact
% to determine if we are running the old or the new class.
\ifundefined{IEEEtransversionmajor}{%
   % This block will be executed only if we are using the old
   % class. All we do is to make sure the V1.3 lengths and commands
   % actually exist so the code won't choke when it
   % doesn't find them.

   % This file doesn't need most of these definitions.
   % However, we'll provide them all in case somebody
   % wants to see what should be executed when compiling
   % a V1.3 or later IEEEtran.cls .tex file with a pre V1.3
   % IEEEtran.cls class file. In such a case, all you have to
   % do is copy this block to the start of your code.
   % However, it won't fix any bugs in the old IEEEtran.cls!
   %
   % **** BACKWARD COMPATIBILITY CODE BLOCK START ****
   \newlength{\IEEEilabelindent}
   \newlength{\IEEEilabelindentA}
   \newlength{\IEEEilabelindentB}
   \newlength{\IEEEelabelindent}
   \newlength{\IEEEdlabelindent}
   \newlength{\labelindent}
   \newlength{\IEEEiednormlabelsep}
   \newlength{\IEEEiedmathlabelsep}
   \newlength{\IEEEiedtopsep}

   \providecommand{\IEEElabelindentfactori}{1.0}
   \providecommand{\IEEElabelindentfactorii}{0.75}
   \providecommand{\IEEElabelindentfactoriii}{0.0}
   \providecommand{\IEEElabelindentfactoriv}{0.0}
   \providecommand{\IEEElabelindentfactorv}{0.0}
   \providecommand{\IEEElabelindentfactorvi}{0.0}
   \providecommand{\labelindentfactor}{1.0}

   \providecommand{\iedlistdecl}{\relax}
   \providecommand{\calcleftmargin}[1]{
                   \setlength{\leftmargin}{#1}
                   \addtolength{\leftmargin}{\labelwidth}
                   \addtolength{\leftmargin}{\labelsep}}
   \providecommand{\setlabelwidth}[1]{
                   \settowidth{\labelwidth}{#1}}
   \providecommand{\usemathlabelsep}{\relax}
   \providecommand{\iedlabeljustifyl}{\relax}
   \providecommand{\iedlabeljustifyc}{\relax}
   \providecommand{\iedlabeljustifyr}{\relax}

   \newif\ifnocalcleftmargin
   \nocalcleftmarginfalse

   \newif\ifnolabelindentfactor
   \nolabelindentfactorfalse

   % in V1.4 of IEEEtran.cls
   \newif\ifcenterfigcaptions
   \centerfigcaptionsfalse

   % we need to provide the old IED environments
   % with a bogus optional argument
   \let\OLDitemize\itemize
   \let\OLDenumerate\enumerate
   \let\OLDdescription\description

   \renewcommand{\itemize}[1][\relax]{\OLDitemize}
   \renewcommand{\enumerate}[1][\relax]{\OLDenumerate}
   \renewcommand{\description}[1][\relax]{\OLDdescription}

   \providecommand{\pubid}[1]{\relax}
   \providecommand{\pubidadjcol}{\relax}
   \providecommand{\specialpapernotice}[1]{\relax}
   \providecommand{\overrideIEEEmargins}{\relax}

   % V1.1 change: use \let instead of \providecommand
   % This prevents LaTeX from hanging if the user ever
   % tried to redefine \PARstart in terms of \CMPARstart
   \let\CMPARstart\PARstart

   \let\OLDappendix\appendix
   \renewcommand{\appendix}[1][\relax]{\OLDappendix}

   \newif\ifuseRomanappendices
   \useRomanappendicestrue

   % V1.2 change: handle the optional biography environment argument
   % (the photo specifier) provided by IEEEtran V1.5 and later.
   % This is tricky because, under the new biography, the SECOND
   % argument is the non-optional one (the biography text).
   \let\OLDbiography\biography
   \let\OLDendbiography\endbiography
   \renewcommand{\biography}[2][\relax]{\OLDbiography{#2}}
   \renewcommand{\endbiography}{\OLDendbiography}
   % **** BACKWARD COMPATIBILITY CODE BLOCK END ****
   % alter the header to show we are using the older class
   \markboth{A Test for IEEEtran.cls--- {\tiny \bfseries
   [Running Older Class]}}{Shell: A Test for IEEEtran.cls}}{
   % END IF OLDER CLASS

   % This block will be executed only if we are running
   % the enhanced class
   % alter the header to show we are using the enhanced class
   \markboth{}%
   {Shell: A Test for IEEEtran.cls}}

% end of conditional

% Uncomment this line to render the big first letter in
% Computer Modern font.
%\renewcommand{\PARstart}[2]{\CMPARstart{#1}{#2}}
% V1.1 change: This would hang if using the older IEEEtran.cls
% in IEEEtest V1.0, it is OK now.
%
%
% Here's how you would do invited papers:
%\specialpapernotice{(Invited Paper)}
% If you are binding copies of work generated with IEEEtran.cls,
% you may want to try:
%\overrideIEEEmargins
% These commands work OK here even though they are not in the preamble.
%(I wanted to put them after the backward compatibility code)

\maketitle
% here's how you get a publisher's ID mark with the new
% IEEEtran.cls.  If you want to use it, don't forget to
% also uncomment the \pubidadjcol command (which must be
% executed in the second text column) around line 434 below
%\pubid{0000--0000/00\$00.00~\copyright~2001 IEEE}

\begin{abstract}
Communications in dispersive direct-sequence code-division
multiple-access channels suffer from intersymbol and
multiple-access interference, which can significantly impair
performance. Joint maximum \textit{a posteriori} probability
equalization and multiuser detection with error control decoding
can be used to mitigate this interference and to achieve the
optimal bit error rate. Unfortunately, such optimal detection
typically requires prohibitive computational complexity. This
problem is addressed in this paper through the development of a
reduced state trellis search detection algorithm, based on
decision feedback from channel decoders. The performance of this
algorithm is analyzed in the large-system limit. This analysis and
simulations show that this reduced complexity algorithm can
exhibit near-optimal performance under moderate signal-to-noise
ratio and attains larger system load capacity than parallel
interference cancellation.
\end{abstract}

% no need for this single page document
% you may have to move \pubidadjcol (if used) if
% these are enabled
%\listoffigures
%\listoftables
%\tableofcontents
\section{Introduction}
Over the past two decades there has been considerable research on
direct sequence code division multiple access (DS-CDMA)
communications, which offers the advantages of soft-capacity
limit, inherent frequency diversity and high data rate
~\cite{Rappaport1999}, and which is the fundamental signaling
technique of third-generation (3G) cellular communications and
other emerging applications.. However, DS-CDMA suffers from
interference, particularly multiple-access interference (MAI),
which is due to the non-orthogonality of different users'
spreading codes, and intersymbol interference (ISI), which is
caused by multipath fading in high-rate systems. It is well known
that mitigation of these types of interference can substantially
improve system performance. In recent years, much progress has
been achieved toward such mitigation through the application of
equalization (EQ) and multiuser detection (MUD).

Equalization can be roughly categorized into three classes. One
class is based on maximum likelihood (ML) detection
~\cite{Proakis2001}, which can be implemented efficiently with the
Viterbi algorithm (VA). A second class is linear equalization
based on some criterion, such as minimum peak distortion or
minimum mean square error (MMSE). The third class is decision
feedback equalization (DFE), in which previously detected symbols
are used to cancel the intersymbol interference
~\cite{Proakis2001}. In multiuser detection, we can find the
counterparts for these three kinds of equalizers. In particular,
ML based multiuser detection and MMSE multiuser detection are both
well known ~\cite{Verdu1998}. As a combination of both techniques,
the problem of joint ISI an MAI mitigation is discussed in
\cite{Beheshti1998}.

In recent years, the turbo principle, namely the iterative
exchange of soft information among different blocks in a
communication system to improve the system performance, has been
applied to equalization and multiuser detection in channel coded
systems, thus resulting in turbo equalization \cite{Laot2001} and
turbo multiuser detection \cite{wang1999}. In these algorithms,
soft decisions from channel decoding are fed back to be used
\textit{a priori} probabilities by a maximum \textit{a posteriori}
probability (MAP) based equalizer or multiuser detector and
enhance the performance iteratively. However, the computational
cost of MAP based detection is prohibitive for large numbers of
users or long delay spread. Thus, it is necessary to reduce the
complexity of such iterative algorithms for practical
applications. For memoryless synchronous multiaccess channels, the
MAP turbo multiuser detection can be simplified to parallel
interference cancellation (PIC) ~\cite{Alex2000}, whose
performance can be enhanced with an MMSE filter ~\cite{wang1999}
or a decorrelating interference canceller ~\cite{Hsu2001}. For
systems with memory, an alternative way to simplify these
detectors is to reduce the number of states by either truncating
the channel memory to a fixed order or eliminating the states with
reliable decisions at the channel decoder. The former strategy has
been used in the equalization of ISI channels
~\cite{DuelHallen1989}~\cite{Eyuboglu1988}~\cite{Muller1996} and
in asynchronous multiuser detection of channel coded CDMA systems
~\cite{Qin2002}, while the latter scheme is of dynamic complexity
and is widely used in iterative decoding algorithms. In speech
recognition ~\cite{Young1996}, this scheme is applied to trim the
acoustic or grammar nodes with low metrics. In early detection
based decoding of parallel turbo codes ~\cite{Frey1998}, the
trellis is simplified by splicing the state with reliable
\textit{a priori} probabilities. Similar strategies have been
applied to the iterative decoding of LDPC codes ~\cite{Fossorier}
and concatenated codes ~\cite{Bokolamulla2003}.

In this paper we study channel coded DS-CDMA systems operating
over frequency selective fading channels. Although joint detection
and decoding can achieve higher channel capacity, we consider only
systems in which detection and decoding are separate due to the
increased feasibility of such systems in practical applications.
In particular, we consider joint equalization and multiuser
detection based on MAP detection with decision feedback (MAP
EQ-MUD) from the decoder. Similarly to MAP turbo equalization or
multiuser detection, the MAP EQ-MUD also suffers from prohibitive
computational cost. In this paper, we first decompose the
multiuser trellis into single-user trellises, thus linearizing the
number of states in terms of the number of users. Then we apply
both the state-reducing techniques in channels with memory, namely
shortening the channel memory to a fixed order and partitioning
the decision feedback of channel decoders into unreliable and
reliable sets of symbols using a simple confidence metric. The
states are constructed with the unreliable set and the
interference from the reliable set is cancelled using the soft
decisions from the decoder. We call this reduced complexity
algorithm \textit{reduced state equalization and multiuser
detection} (RS EQ-MUD).

This paper is organized as follows. Section II introduces the
system model, in which the signal model and decoder are explained.
The optimal MAP EQ-MUD algorithm is described in Section III,
while Section IV is focused on developing the RS EQ-MUD algorithm.
An asymptotic analysis of the system performance is carried out in
Section V, and corresponding numerical results are given in
Section VI. Final conclusions are drawn in Section VII.

\section{System Model}
\subsection{Signal model}
For a channel-coded DS-CDMA system, let $K$ denote the number of
active users, $N$ the spreading gain and $M$ the coded symbol
block length. Denote the symbol period and chip period by $T_s$
and $T_c$ respectively, and note that $T_s=NT_c$. For user $k$,
the binary phase-shift keying (BPSK) modulated continuous-time
signal at the transmitter is given by
\begin{eqnarray}
  \tilde{r}_k(t)&=&\sum_{i=1}^M b_k(i)s_k(t-iT_s)        \nonumber \\
        &=&\sum_{i=1}^{M} b_k(i) \sum_{n=1}^Ns_{k}^{(i)}(n)\Psi
        (t-iT_s-nT_c),
\end{eqnarray}
where $b_k(i)$ is the \textit{i}-th (binary) channel coded symbol
sent by user $k$, $\Psi(t)$ is the chip waveform, and
$s_{k}^{(i)}(n)$ is the normalized binary spreading code of user
\textit{k}, which satisfies
$\left|s_k^{(i)}(n)\right|=\frac{1}{\sqrt{N}}$. The superscript
$i$ in $s_{k}^{(i)}(n)$ implies that the spreading code varies
with the symbol period, since long (aperiodic) codes are
considered in this paper.

The signal (1) passes through a frequency selective fading channel
whose impulse response is $\tilde{h}_k(t)$, and the channel output
is the convolution of the input signal and the channel response:
\begin{eqnarray}
  r_k(t)&=&\tilde{r}_k(t)\star \tilde{h}_k(t);        \nonumber \\
        &=&\sum_{i=1}^{M}b_k(i)\sum_{n=1}^Ns_k^{(i)}(n)g_k(t-iT_s-nT_c),
\end{eqnarray}
where
\begin{eqnarray}
  g_k(t-iT_s-nT_c)=\Psi(t-iT_s-nT_c)\star \tilde{h}_k(t).
\end{eqnarray}

Assuming synchronous transmission among users \footnote{Note that
this assumption of synchronous transmission does not limit the
generality of this model, since delay offsets between users can be
incorporated into the channel impulse response $\tilde{h}_1(l)$,
..., $\tilde{h}_K(l)$.}, the signal at the receiver is given by
\begin{eqnarray}
  r(t)&=&\sum_{k=1}^Kr_k(t)     \nonumber \\
        &=&\sum_{k=1}^K\sum_{i=1}^{M}b_k(i)\sum_{n=1}^Ns_k^{(i)}(n)g_k(t-iT_s-nT_c).
\end{eqnarray}
The received signal is sampled at the chip rate $\frac{1}{T_c}$,
and the corresponding discrete output of the sampler at chip
period $l$ is
\begin{eqnarray}
  y(l)=r(lT_c)&=&\sum_{k=1}^K\sum_{i=1}^{M}b_k(i)\sum_{n=1}^Ns_k^{(i)}(n)g_k((l-n-iN)T_c)     \nonumber \\
        &=&\sum_{k=1}^K\sum_{i=1}^{M}b_k(i)h_k^{(i)}(l-iN),
\end{eqnarray}
where
\begin{eqnarray}
  h_k^{(i)}(l)=s_k^{(i)}(l)\star g_k(lT_c).
\end{eqnarray}
Suppose the support of $h_k^{(i)}(l)$ is $(0,(L-1)N)$, where $L$
is the dispersion length. We can simplify (5) to a vector form.
(For notational convenience, we henceforth use $t$ to designate
discrete time advancing at the symbol rate.) This vector is given
by \footnote{$T$ denotes transposition, $H$ denotes conjugate
transposition and * denotes conjugation.}
$$\textbf{y}(t)=\left(y(tN),y(tN+1),...,y((t+1)N-1)\right)^T,$$
and
$$\textbf{h}^{(i)}_k(j)=\left(h^{(i)}_k(jN),h^{(i)}_k(jN+1),...,h^{(i)}_k((j+1)N-1))\right)^T.$$
Using the vector form and considering the thermal noise at the
receiver, we write
\begin{eqnarray}
  \textbf{y}(t)=\sum_{k=1}^K\sum_{i=t-L+1}^{t}b_k(i)\textbf{h}_k^{(i)}(t-i)+\textbf{n}(t),\qquad
  t=1,2,..., M
\end{eqnarray}
where $\textbf{n}(t)$ is additive white Gaussian noise (AWGN) that
satisfies
$E\left\{\textbf{n}(t)\textbf{n}(t)^H\right\}=\sigma_n^2I_{N\times
N }$. The development in the remaining of this paper will be based
on this discrete model (7).
\subsection{Equivalent spreading code}
We term $\textbf{h}_k^{(i)}(t-i)$ the equivalent spreading code of
the $i$-th symbol of user $k$ in the $t$-th $(i\leq t)$ symbol
period. In order to explore the properties of the equivalent
spreading code, we need to place some assumptions on the discrete
channel response $g_k(lT_c)$ in (6):
\begin{itemize}
\item Causality. $g_k(lT_c)=0$ when $l<0$.
\item Normality. We assume that $\left\{g_k(lT_c)\right\}_l$ is a complex-valued
Gaussian random sequence with zero mean and exponentially decaying
variance
\begin{eqnarray}
  \sigma_k^2(l)=\frac{\lambda_ke^{-\lambda_k\frac{l}{N}}}{N},
\end{eqnarray}
where $\lambda_k$ is the decay factor and satisfies
$e^{-\lambda_kL}\approx 0$ and $\lim_{N\rightarrow
\infty}\sum_{l=1}^\infty \sigma_k^2(l)=1$.
\item Independence. We assume that the scattering caused by fading is
uncorrelated, which means
\begin{eqnarray}
  E\left\{g_k(lT_c)g_n(jT_c)^*\right\}=0, \qquad \mbox{if $l\neq j$ or $k\neq n$}.
\end{eqnarray}
Thus $g_k(lT_c)$ and $g_n(jT_c)$ are independent since they are
jointly Gaussian.
\end{itemize}

In non-dispersive multiple-access channels, the cross correlation
of the spreading codes plays a key role in the system performance.
Thus we need to discuss the cross correlation between the
equivalent spreading codes. Define the equivalent cross
correlation between the equivalent spreading codes of user $k$ and
user $n$ to be
\begin{eqnarray}
  \rho_{kn}^{t\tau}(i,j)=\left(\textbf{h}_k^{(t)}(i)\right)^H\textbf{h}_n^{\left(\tau\right)}(j).
\end{eqnarray}
Note that $\rho_{kn}^{t\tau}(i,j)$ is a random variable since
$\textbf{h}_k^t(i)$ and $\textbf{h}_n^{\tau}(j)$ are random.

\subsection{Channel decoder}
The diagram of the system discussed in this paper is given in Fig.
1. At the transmitter, the information symbols are encoded,
interleaved with an infinite length interleaver and spread. The
BCJR algorithm ~\cite{BCJR} is used in the channel decoder, which
follows a deinterleaver, to obtain the \textit{a posteriori}
probability $P(b_k(t)=b|\textbf{y}_1^M)$, where
$\textbf{y}_1^M=\{\textbf{y}(t)|t=1,...,M\}$. This probability can
be decomposed into two parts,
$$
  P(b_k(t)=b|\textbf{y}_1^M)\propto
  P(b_k(t)=b|\textbf{y}(t))\zeta_k^t(b),
$$
where $\zeta_k^t(b)$ is the extrinsic information about $b_k(t)$
from the other coded symbols. We use the extrinsic information to
construct the soft decision feedback:
\begin{eqnarray}
  \hat{b}_k(t)=2\zeta_k^t(1)-1,
\end{eqnarray}
which is used to cancel both the ISI and MAI. The estimation error
is denoted by $\Delta b_k(t)=b_k(t)-\hat{b}_k(t)$; its expectation
is zero due to symmetry and its variance can be obtained by
simulation.

\section{Optimal MAP Turbo EQ-MUD}
In this section we develop the BCJR algorithm ~\cite{BCJR} for MAP
EQ-MUD in a way similar to that used for turbo multiuser detection
in ~\cite{wang1999}.

First we define the state at symbol period $t$ as the set of all
symbols of all users from symbol period $t-L+2$ to $t$:
\begin{eqnarray}
\textbf{S}_t=\left\{b_k(l)|l=t-L+2,...,t,\mbox{
}k=1,...,K\right\}.
\end{eqnarray}
Thus for each symbol period we have $2^{(L-1)K}$ states with which
to construct a trellis. We call state $\textbf{m}$ and
$\textbf{m}'$ \textit{compatible} when the state can transit from
$\textbf{m}$ to $\textbf{m}'$ and denote this condition by
$\textbf{m}\Rightarrow \textbf{m}'$.

Next we define some intermediate variables \cite{BCJR}:
\begin{itemize}
\item forward probability:
$\alpha_t(\textbf{m})=P(\textbf{S}_t=\textbf{m},\textbf{y}^t_1)$;
\item backward probability:
$\beta_t(\textbf{m})=P(\textbf{y}^M_{t+1}|\textbf{S}_t=\textbf{m})$;
\item transition probability:
$\gamma_t(\textbf{m}',\textbf{m})=P(\textbf{y}(t),\textbf{S}_t=\textbf{m}|\textbf{S}_{t-1}=\textbf{m}')$.
\end{itemize}
The forward and backward probabilities can be computed recursively
via the equations
\begin{eqnarray}
   \alpha_t(\textbf{m})&=&\sum_{\textbf{m}'}\alpha_{t-1}(\textbf{m}')\gamma_t(\textbf{m}',\textbf{m}),
\end{eqnarray}
and
\begin{eqnarray}
   \beta_t(\textbf{m})&=&\sum_{\textbf{m}'}\beta_{t+1}(\textbf{m}')\gamma_{t+1}(\textbf{m},\textbf{m}').
\end{eqnarray}
Invoking Bayes' formula, we can rewrite the transition probability
as
$\gamma_t(\textbf{m},\textbf{m}')=P(\textbf{y}(t)|\textbf{S}_t=\textbf{m},\textbf{S}_{t-1}=\textbf{m}')P(\textbf{S}_t=\textbf{m}|\textbf{S}_{t-1}=\textbf{m}')$.
Since we assume that the interleaver has infinite length, the
symbols of different users and different symbol periods are
independent. Thus we have
\begin{eqnarray}
P(\textbf{S}_t=\textbf{m}|\textbf{S}_{t-1}=\textbf{m}')&=&P(b_1(t)=\tilde{b}_1,...,b_K(t)=\tilde{b}_K|\tilde{b}_1,...,\tilde{b}_K\in \textbf{m} )\nonumber \\
                     &=&\prod_{k=1}^KP\left(b_k(t)=\tilde{b}_k|\tilde{b}_k\in \textbf{m}\right)\nonumber \\
                     &=&\prod_{\tilde{b}_k\in \textbf{m}}\zeta_k^t\left(\tilde{b}_k\right)\nonumber \\
                     &\triangleq&\Lambda_t(\textbf{m}),
\end{eqnarray}
which can be regarded as the $\textit{a priori}$ probabilities of
the symbols at symbol period $t$. When no $\textit{a priori}$
information is available, we assume the symbols to be uniformly
distributed. Since the channel decoder can provide the extrinsic
information from the other coded symbols, we can feed the soft
outputs of the decoder back as the $\textit{a priori}$
probability. Hence
\begin{eqnarray}
 \gamma_t(\textbf{m}',\textbf{m}) = \left\{ \begin{array}{lll}
        \Lambda_t(\textbf{m})\frac{1}{\sqrt{2\pi\sigma_n^2}}\exp\left(-\frac{\left\|\textbf{y}(t)-\hat{\textbf{y}}_{\textbf{m},\textbf{m}'}(t)\right\|^2}{2\sigma_n^2}\right)  & \mbox{if $\textbf{m}' \Rightarrow \textbf{m}$};\\
        0 & \mbox{if $\textbf{m}' \nRightarrow \textbf{m}$}.\end{array} \right.
        \nonumber,
\end{eqnarray}
where $\hat{\textbf{y}}_{\textbf{m},\textbf{m}'}(t)$ is the
estimated received signal due to the states $\textbf{m}$ and
$\textbf{m}'$:
\begin{eqnarray}
 \hat{\textbf{y}}_{\textbf{m},\textbf{m}'}(t)=\sum_{k=1}^K\sum_{i=t-L+1}^{t}\tilde{b}_k(i)\textbf{h}_k^{(t)}(t-i),\qquad
 \tilde{b}_k(i)\in \textbf{m}'\cup \textbf{m}.
\end{eqnarray}

When the forward and backward probabilities are available, we can
compute the joint probability
\begin{eqnarray}
   P\left(\textbf{S}_t=\textbf{m},\textbf{y}_1^M\right)=\alpha_{t}(\textbf{m})\beta_t(\textbf{m}),
\end{eqnarray}
and then compute the \textit{a posteriori} probability:
\begin{eqnarray}
   P\left(b_k(t)=\tilde{b}_k|\textbf{y}_1^M\right)\propto \sum_{\tilde{b}_k\in
   \textbf{m}}P\left(\textbf{S}_{t}=\textbf{m},\textbf{y}_1^M\right),
\end{eqnarray}
where $\propto$ denotes proportionality. The appropriate
normalization is
$$
P\left(b_k(t)=1|\textbf{y}_1^M\right)+P\left(b_k(t)=-1|\textbf{y}_1^M\right)=1.
$$
To avoid the reuse of the extrinsic information, we need to cancel
the $\textit{a priori}$ probability. Therefore the soft input to
the decoder, denoted by $\mathcal{L}_k(t)$, for symbol $t$ of user
$k$ is normalized by the soft decision feedback:
\begin{eqnarray}
   \mathcal{L}_k(t)=\frac{\left[\frac{P\left(b_k(t)=1|\textbf{y}_1^M\right)}{P\left(b_k(t)=-1|\textbf{y}_1^M\right)}\right]}{\left[\frac{\zeta(b_k(t)=1)}{\zeta(b_k(t)=-1)}\right]}.
\end{eqnarray}

\section{Reduced State EQ-MUD}
\subsection{Independence assumption and trellis decomposition}
Facing the same problem as either equalization or multiuser
detection, the optimal MAP EQ-MUD suffers from prohibitive
computational complexity because the number of states increases
exponentially with the number of users $K$ and the dispersion
length $L$. Thus, the optimal MAP EQ-MUD is primarily of
theoretical value and cannot be implemented for many practical
applications.

In this paper we mitigate the complexity of the MAP EQ-MUD, by
decomposing its trellis into single-user sub-trellises, thus
linearizing the number of states with respect to the number of
users. For user $k$, we define the sub-state at symbol period $t$
as
$$
S_t^k=\{b_k(t-L+2),...,b_k(t)\}.
$$
The sub-states belonging to one user construct a single-user
sub-trellis. Then any state defined in (12) is the union of the
corresponding sub-states of all users. In order to distinguish
sub-state and state, we use non-bold fonts to designate the
sub-states in the rest of this paper. Similar to Section III, we
define the forward and backward probabilities for the sub-state of
user $k$:
\begin{eqnarray}
    \alpha_t^k(m)&=&P\left(S_t^k=m,\textbf{y}^t_1\right),
\end{eqnarray}
and
\begin{eqnarray}
    \beta_t^k(m)&=&P\left(S_t^k=m,\textbf{y}^M_{t+1}\right).
\end{eqnarray}
Compared with the analogous definition of Section III, the
definition of the backward probability $\beta_t^k(m)$ is slightly
different here; this will facilitate the application of Assumption
IV.1, which follows immediately, on the backward probability. If
the BCJR algorithm can be confined to each sub-trellis, the number
of sub-states will be reduced to $K2^{L-1}$. However, the forward
and backward probabilities defined in Section III involve joint
distributions of different users' symbols; thus the sub-trellis
searches for different users are coupled, which prohibits the
exact decomposition into single-user trellises.

However, we can approximate the joint distribution with the
product of marginal distributions, which results in the following
assumption.
\begin{assum}
For any state $\textbf{S}_t$ defined in (12) which is the union of
the correspongding sub-states,
$\textbf{S}_t=\bigcup_{k=1}^KS_t^k$, we have
\begin{eqnarray}
   P\left(\textbf{S}_t|\textbf{y}_1^t\right)&=&\prod_{k=1}^KP\left(S_t^k|\textbf{y}_1^t\right),
\end{eqnarray}
and
\begin{eqnarray}
   P\left(\textbf{S}_t|\textbf{y}_{t+1}^M\right)&=&\prod_{k=1}^KP\left(S_t^k|\textbf{y}_{t+1}^M\right).
\end{eqnarray}
\end{assum}

This assumption is based on the fact that
$P\left(S_t^k|\textbf{y}_1^t\right)$ is concentrated around 0 and
1, provided that the noise power is small enough. It can be
validated by the simulation results in Section VI, which state
that, with Assumption IV.1, the reduced complexity algorithm in
this paper can achieve near optimal performance in moderate energy
region. Under this assumption, which is only an approximation, we
can decompose the forward and backward probabilities of a state
into the product of the probabilities of the corresponding
sub-states:
\begin{eqnarray}
    \alpha_t(\textbf{S}_t)&=&\frac{\prod_{k=1}^K\alpha_t^k(S_t^k)}{\left(P(\textbf{y}_1^t)\right)^{K-1}},
\end{eqnarray}
and
\begin{eqnarray}
    \beta_t(\textbf{S}_t)&=&\frac{\prod_{k=1}^K\beta_t^k(S_t^k)}{\left(P(\textbf{y}_{t+1}^M)\right)^{K-1}}.
\end{eqnarray}
We can neglect the common factors
$\left(P(\textbf{y}_1^t)\right)^{K-1}$ and
$\left(P(\textbf{y}_{t+1}^M)\right)^{K-1}$, which do not affect
the final result.

With this assumption, we can develop recursive formulas similar to
(13) and (14) with respect to the sub-states. In particular, for
the forward probability, we have
\begin{eqnarray}
    \alpha_t^k(m)&=&P\left(S_t^k=m,\textbf{y}^t_1\right)  \nonumber \\
                 &=&\sum_{\textbf{m}'}\sum_{m\in\textbf{m}''}P\left(\textbf{S}_t=\textbf{m}'',\textbf{S}_{t-1}=\textbf{m}',\textbf{y}_1^{t}\right)\nonumber\\
                 &=&\sum_{\textbf{m}'}\sum_{m\in\textbf{m}''}P\left(\textbf{y}_1^{t}|\textbf{S}_t=\textbf{m}'',\textbf{S}_{t-1}=\textbf{m}'\right)P\left(\textbf{S}_t=\textbf{m}''|\textbf{S}_{t-1}=\textbf{m}'\right)P\left(\textbf{S}_{t-1}=\textbf{m}'\right)\nonumber                 \\
                 &=&\sum_{\textbf{m}'}\sum_{m\in\textbf{m}''}P\left(\textbf{y}_1^{t-1},\textbf{S}_{t-1}=\textbf{m}'\right)P\left(\textbf{y}(t)|\textbf{S}_t=\textbf{m}'',\textbf{S}_{t-1}=\textbf{m}'\right)P\left(\textbf{S}_t=\textbf{m}''|\textbf{S}_{t-1}=\textbf{m}'\right)\nonumber                 \\
                 &=&\sum_{\textbf{m}'}\left(\left(\prod_{i=1, m_i\in
                 \textbf{m}'}^K\alpha_{t-1}^i(m_i)\right)\lambda_t^k(\textbf{m}',m)\right),
\end{eqnarray}
where $\lambda_t^k(\textbf{m}',m)$ is the probability of
transition from state $\textbf{m}'$ to sub-state $m$.
\begin{eqnarray}
  \lambda_t^k(\textbf{m}',m)&\triangleq&\sum_{m\in\textbf{m}''}P\left(\textbf{y}(t),\textbf{S}_t=\textbf{m}''|\textbf{S}_{t-1}=\textbf{m}'\right)  \nonumber \\
                            &=&\sum_{m\in\textbf{m}''}P\left(\textbf{y}(t)|\textbf{S}_t=\textbf{m}'',\textbf{S}_{t-1}=\textbf{m}'\right)\Lambda_t(\textbf{m}'')\nonumber                            \\
                            &=&\sum_{m\in\textbf{m}''}\Lambda_t(\textbf{m}'')\frac{1}{\sqrt{2\pi\sigma_n^2}}\exp\left(-\frac{\|\textbf{y}(t)-\hat{\textbf{w}}_{\textbf{m}'',\textbf{m}'}(t)\|^2}{2\sigma_n^2}\right),
\end{eqnarray}
where $\hat{\textbf{w}}_{\textbf{m}'',\textbf{m}'}(t)$ is the
estimated received signal due to the states $\textbf{m}'$ and
$\textbf{m}''$:
\begin{eqnarray}
 \hat{\textbf{w}}_{\textbf{m}'',\textbf{m}'}(t)=\sum_{n=1}^K\sum_{i=t-L+1}^{t}\tilde{b}_n(i)\textbf{h}_n^{(i)}(t-i),
\end{eqnarray}
where $\tilde{b}_n(i)\in \textbf{m}'\cup \textbf{m}''$.

For the backward probability $\beta_t^k(m)$, we can obtain a
similar recursive formula with the same manipulations as the
forward probability:
\begin{eqnarray}
  \beta_t^k(m)&=&\sum_{m\in\textbf{m}''}\sum_{\textbf{m}'}P\left(\textbf{y}_{t+2}^M,\textbf{S}_{t+1}=\textbf{m}'\right)P\left(\textbf{y}(t+1)|\textbf{S}_t=\textbf{m}'',\textbf{S}_{t+1}=\textbf{m}'\right)P\left(\textbf{S}_t=\textbf{m}''|\textbf{S}_{t+1}=\textbf{m}'\right)\nonumber  \\
              &=&\sum_{\textbf{m}'}\left(\left(\prod_{i=1, m_i\in \textbf{m}'}^K\beta_{t+1}^i(m_i)\right)\phi_{t+1}^k(m,\textbf{m}')\right),
\end{eqnarray}
where $\phi_{t+1}^k(m,\textbf{m}')$ is the probability of
transition from sub-state $m$ to state $\textbf{m}'$. With the
same manipulation as $\lambda_t^k(\textbf{m}',m)$, we have
\begin{eqnarray}
  \phi_{t+1}^k(m,\textbf{m}')=\sum_{m\in\textbf{m}''}\Lambda_{t-L+2}(\textbf{m}'')\frac{1}{\sqrt{2\pi\sigma_n^2}}\exp\left(-\frac{\|\textbf{r}(t+1)-\hat{\textbf{v}}_{\textbf{m}'',\textbf{m}'}(t+1)\|^2}{2\sigma_n^2}\right),
\end{eqnarray}
where $\hat{\textbf{v}}_{\textbf{m}'',\textbf{m}'}(t+1)$ is the
estimated received signal due to the states $\textbf{m}'$ and
$\textbf{m}''$:
\begin{eqnarray}
 \hat{\textbf{v}}_{\textbf{m}'',\textbf{m}'}(t+1)&=&\sum_{n=1}^K\sum_{i=t-L+2}^{t+1}\tilde{b}_n(i)\textbf{h}_n^{(i)}(t+1-i),
\end{eqnarray}
where $\tilde{b}_n(i)\in \textbf{m}'\cup\textbf{m}''$.

When the forward and backward probabilities are available, we can
compute the joint probability in a way similar to (17):
\begin{eqnarray}
P\left(S_t^k=m,\textbf{y}_1^M\right)=\frac{\alpha_{t}^k(m)\beta_{t}^k(m)}{P(S_t^k=m)}.
\end{eqnarray}
And the likelihood ratio for input to the decoder can be computed
with (18) and (19).

\subsection{Reduced state trellis}
With the above independence assumption, we have reduced the number
of sub-states to $K2^{L-1}$. However, the BCJR algorithm of each
sub-trellis is still coupled to the other sub-trellises because of
the existence of MAI. In (26) and (29) the number of terms in the
double summations increases exponentially with $K$. Thus the
computational complexity of the decomposed trellis still remains
prohibitive for practical implementations, and we need reduce
further the number of sub-states. We can accomplish this with the
aid of the decision feedback from the decoders. The underlying
philosophy is to partition the decision feedbacks into reliable
and unreliable sets, denoted by $\textbf{R}$ and
$\bar{\textbf{R}}$. The trellis is constructed with the symbols in
the unreliable set while the symbols in the reliable set are
considered to be known and are directly cancelled from the
original signal.

Suppose that the decision errors of the decoders are $\{\Delta
b_k(t)\}_{\begin{scriptsize}\begin{array}{ll}k=1,2,...,K\\
t=1,2,...,M\end{array}\end{scriptsize}}$. We sort the absolute
values of these errors and select the largest $\hat{K}$ ones to
form the unreliable set $\bar{\textbf{R}}$ while the remaining
symbols comprise the reliable set $\textbf{R}$. We define
$\kappa=\frac{\hat{K}}{KM}$ as the normalized search width, which
is also the probability that a decision feedback symbol is
unreliable.

When we are constructing the sub-states at symbol period $t$, we
consider only the symbols from $t-L_1+1$ to $t$ where $L_1$ is an
integer with $1\leq L_1\leq L$, thus shortening the channel memory
in a way similar to
~\cite{DuelHallen1989}~\cite{Eyuboglu1988}~\cite{Muller1996}.
$L_1$ can be selected so that the decision feedback errors of the
symbols earlier than $L_1$ compose only a small portion of the
interference due to the exponential decay of the channel. With the
unreliable set $\bar{\textbf{R}}$ and notation
$\bar{\textbf{R}}(t)=\{(k,i)\in \bar{\textbf{R}}, t-L_1+1\leq i<
t, 1\leq k\leq K\}$, we define the sub-states of user $k$ at
symbol period $t$:
$$
S_t^k=\left\{b_k(i)|(k,i)\in \bar{\textbf{R}}(t) \mbox{ or } i =
t\right\}.
$$
Observe that after reducing the sub-states, the numbers of
sub-states of different users may be different and depend on the
selection of the unreliable set. Thus the total number of
sub-states at symbol period $t$ is $\sum_{k=1}^K2^{\mu_k+1}$,
where $\mu_k$ is the number of unreliable symbols of user $k$ from
symbol period $t-L_1$ to $t-1$ and has the constraint
$\sum_{k=1}^K\mu_k=\mbox{the size of set } \bar{\textbf{R}}(t)$.
Assuming that $\mu_k$ is Bernoulli distributed, the mean value of
the number of sub-states at symbol period $t$ is
$K\sum_{m=0}^{L_1-1}{m\choose L_1-1
}\kappa^m(1-\kappa)^{L_1-1-m}2^{m+1}$.

The BCJR algorithm for decoding the reduced sub-state trellis is
also confined to the unreliable set. The transition probability is
nonzero only when $\textbf{m}'\Rightarrow m$ and
$\textbf{m}'\subset \bar{\textbf{R}}(t)$ for (26) and when $m
\Rightarrow\textbf{m}'$ and $\textbf{m}'\subset
\bar{\textbf{R}}(t+1)$ for (29). The recursive formulas are
rewritten as follows:
\begin{eqnarray}
    \alpha_t^k(m)&=&\sum_{\textbf{m}'\subset \bar{\textbf{R}}(t)}\left(\left(\prod_{i=1, m_i\in \textbf{m}'}^K\alpha_{t-1}^i(m_i)\right)\lambda_t^k(\textbf{m}',m)\right),
\end{eqnarray}
and
\begin{eqnarray}
    \beta_t^k(m)&=&\sum_{\textbf{m}'\subset \bar{\textbf{R}}(t+1)}\left(\left(\prod_{i=1,m_i\in
    \textbf{m}'}^K\beta_{t+1}^i(m_i)\right)\phi_{t+1}^k(m,\textbf{m}')\right).
\end{eqnarray}
And we rewrite (28) and (31) as follows:
\begin{eqnarray}
 \hat{\textbf{w}}_{\textbf{m}'',\textbf{m}'}(t)=\sum_{\begin{scriptsize}\begin{array}{l}\tilde{b}_n(i)\in \textbf{m}'\cup\textbf{m}''\end{array}\end{scriptsize}}\tilde{b}_n(i)\textbf{h}_n^{(i)}(t-i)
                                             +\sum_{\begin{scriptsize}(n,i)\notin\textbf{m}'\cup\textbf{m}''\end{scriptsize}}\hat{b}_n(i)\textbf{h}_n^{(i)}(t-i),
\end{eqnarray}
and
\begin{eqnarray}
 \hat{\textbf{v}}_{\textbf{m}'',\textbf{m}'}(t+1)=\sum_{\begin{scriptsize}\begin{array}{l}\tilde{b}_n(i)\in \textbf{m}'\cup\textbf{m}''\end{array}\end{scriptsize}}\tilde{b}_n(i)\textbf{h}_n^{(i)}(t+1-i)
                                             +\sum_{\begin{scriptsize}(n,i)\notin\textbf{m}'\cup\textbf{m}''\end{scriptsize}}\hat{b}_n(i)\textbf{h}_n^{(i)}(t+1-i),
\end{eqnarray}
where $\hat{b}_n(i)$ is the reliable decision feedback about
symbol $b_n(i)$ from the decoder.

Figure 2 illustrates the search of the sub-trellis with $K=2$ and
$L_1=2$. The state transitions from symbol period $t-1$ to $t$ are
labelled with arrows in the figure. Since $\hat{b}_2(t-1)$ is
reliable, the corresponding interference is cancelled directly.
Thus we consider only the transitions from $S_{t-1}^1$ to $S_t^1$
and $S_t^2$ since $\hat{b}_1(t-1)$ is unreliable.

Since the decision feedback errors are unknown to the receiver, we
cannot use $\{\Delta b_k(t)\}$ to construct the unreliable set and
it can only provide an upper bound for the performance. It is easy
to show that the larger the absolute value of $\hat{b}_k(t)$ is,
the more reliable the soft decision feedback is, when only the
soft decision feedbacks are available. Thus we call $|\hat{b}|$
the \textit{confidence metric} for symbol $b$ and, use it to
construct the unreliable set.

\subsection{Distribution of the unreliable set}
For asymptotic analysis of system performance in Section V, we
need to obtain the asymptotic distribution of decision feedback
error in the unreliable set. Suppose the probability density
function (pdf) of $|\hat{b}|$ is $f_{|\hat{b}|}(x)$ and the
cumulative distribution function (cdf) is $F_{|\hat{b}|}(x)$. Then
the pdf of the $k$-th smallest $|\hat{b}|$ out of $n$
independently and identically distributed (i.i.d.) $|\hat{b}|$'s
is
\begin{eqnarray}
   f_{|\hat{b}|}^{(k)}(x)=\frac{n!}{(k-1)!(n-k)!}\left(F_{|\hat{b}|}(x)\right)^{k-1}\left(1-F_{|\hat{b}|}(x)\right)^{n-k}f_{|\hat{b}|}(x).
\end{eqnarray}
As $K\rightarrow\infty$, $f_{|\hat{b}|}^{(k)}(x)$ converges to a
simplified asymptotic expression according to the following lemma
~\cite{Shamai2002}.
\begin{lem}
Let $f^{(k)}(x)$ be the pdf of the $k$-th smallest element out of
$n$ i.i.d. random variables whose pdf's are $f(x)$ and cdf's are
$F(x)$. As $n\rightarrow\infty$,
$f^{(k)}(x)\rightarrow\delta(x-F^{-1}(\frac{k}{n}))$, where
$\delta(.)$ is the Dirac delta function.
\end{lem}

With Lemma IV.2, we can prove the following theorem, which
provides sufficient and necessary condition for unreliable
decision feedback in asymptotic sense.
\begin{thm}
As $K\rightarrow \infty$, $(n,i) \in \bar{\textbf{R}}$ if and only
if $|\hat{b}_n(i)|< F_{|\hat{b}|}^{-1}(\kappa)$, almost surely.
\end{thm}
\begin{proof}
When $|\hat{b}_n(i)|=x< F_{|\hat{b}|}^{-1}(\kappa)$,
\begin{small}
\begin{eqnarray}
  P((n,i)\in \bar{\textbf{R}})&=&\sum_{p=1}^{\kappa KM}P\left(|\hat{b}_n(i)|\mbox{ is } \mbox{the p-th smallest}||\hat{b}_n(i)|=x\right)\nonumber \\
                             &=&\sum_{p=1}^{\kappa KM}\frac{P\left(|\hat{b}_n(i)|=x||\hat{b}_n(i)|\mbox{ is } \mbox{the p-th smallest}\right)P\left(|\hat{b}_n(i)|\mbox{ is } \mbox{the p-th smallest}\right)}{P\left(|\hat{b}_n(i)|=x\right)} \nonumber\\
                             &=&\sum_{p=1}^{\kappa KM}\frac{f_{|\hat{b}|}^{(p)}(x)}{KMf_{|\hat{b}|}(x)}\nonumber \\
                             &\rightarrow&\int_0^{\kappa}\frac{\delta\left(x-F_{|\hat{b}|}^{-1}(\tau)\right)}{f_{|\hat{b}|}(x)}d\tau, \qquad \mbox{as }K\rightarrow\infty \nonumber \\
                             &=&1.\nonumber
\end{eqnarray}
\end{small}
When $(n,i) \in \bar{\textbf{R}}$,
\begin{small}
\begin{eqnarray}
P\left(x\geq
F_{|\hat{b}|}^{-1}(\kappa)|(n,i)\in\bar{\textbf{R}}\right)&=&\frac{\sum_{p=1}^{\kappa
KM}P\left(x<F_{|\hat{b}|}^{-1}(\kappa)||\hat{b}_n(i)|\mbox{ is }
\mbox{the p-th smallest}\right)P\left(|\hat{b}_n(i)|\mbox{ is }
\mbox{the p-th
smallest}\right)}{P\left((n,i)\in\bar{\textbf{R}}\right)}\nonumber \\
                 &=&\frac{\sum_{p=1}^{\kappa KM}\int_{F_{|\hat{b}|}^{-1}(\kappa)}^1f_{|\Delta_b|}^{(p)}(x)dx}{P((n,i)\in\bar{\textbf{R}})KM}\nonumber\\
                 &\rightarrow&\frac{1}{P((n,i)\in\bar{\textbf{R}})}\int_0^{\kappa}\int_{F_{|\hat{b}|}^{-1}(\kappa)}^1\delta(x-F_{|\hat{b}|}^{-1}(\tau))dxd\tau  \nonumber \\
                 &=&0\nonumber.
\end{eqnarray}
\end{small}
This completes the proof.
\end{proof}

With Theorem IV.3 we can obtain the asymptotic pdf
$f_{|\hat{b}|}^{\bar{\textbf{R}}}(x)$ of the unreliable set
constructed by the confidence metric,
$$
  f_{|\hat{b}|}^{\bar{\textbf{R}}}(x)=Cf_{|\hat{b}|}(x)u(-x+\tilde{\kappa}),
$$
where $u(x)$ is the unit step function,
$\tilde{\kappa}=F_{|\hat{b}|}^{-1}(\kappa)$ and
$C=\frac{1}{\int_0^{\tilde{\kappa}}f_{|\hat{b}|}(x)}$ is a
normalizing constant.

However, what we are really interested in is the distribution of
the decision feedback error in the unreliable set. Let $0\leq
x\leq 2$, then
\begin{eqnarray}
            P\left(|\Delta b|<x|\bar{\textbf{R}}\right)&=&P\left(|\Delta b|<x\mid|\hat{b}|<\tilde{\kappa}\right)\nonumber   \\
                                              &=&\frac{P\left(|\Delta b|<x,|\hat{b}|<\tilde{\kappa}\right)}{P(\left(|\hat{b}|<\tilde{\kappa}\right)}\nonumber    \nonumber  \\
                                              &=&\frac{1}{2\kappa}\left(P\left(0<\Delta b<x,|1-\Delta b|<\tilde{\kappa}\right)+P\left(-x<\Delta b<0,|-1-\Delta b|<\tilde{\kappa}\right)\right)\nonumber          \\
                                              &=&\frac{1}{2\kappa}\left(P(1-\tilde{\kappa}<\Delta b<\min(x,1+\tilde{\kappa}))+P(\max(-x,-1-\tilde{\kappa})<\Delta b<-1+\tilde{\kappa})\right)\nonumber\\
                                              &=&\left\{\begin{array}{lll} \frac{1}{\kappa}P\left(1-\tilde{\kappa}<|\Delta b|<x\right), \qquad &\mbox{if $1-\tilde{\kappa}<x<1+\tilde{\kappa}$}\\
                                                          1,  \qquad &\mbox{if $x\geq 1+\tilde{\kappa}$}\\
                                                          0,  \qquad &\mbox{if $x\leq 1-\tilde{\kappa}$}\\
                                                           \end{array}\right.\nonumber.
\end{eqnarray}
Thus the asymptotic pdf of the decision error in
$\bar{\textbf{R}}$ is
\begin{eqnarray}
   f_{|\Delta b|}^{\bar{\textbf{R}}}(x)=\frac{1}{\kappa}f_{|\Delta
   b|}(x)u(x-1+\tilde{\kappa})u(-x+\tilde{\kappa}+1).
\end{eqnarray}
With the same manipulation, we can obtain the asymptotic pdf of
the errors in the reliable set $\textbf{R}$:
\begin{eqnarray}
   f_{|\Delta b|}^{\textbf{R}}(x)=\frac{1}{1-\kappa}f_{|\Delta
   b|}(x)\left(u(x-1-\tilde{\kappa})+u(-x-\tilde{\kappa}+1)\right).
\end{eqnarray}

\section{Asymptotic Performance Analysis}
\subsection{Asymptotic performance analysis}
To evaluate the system performance of RS EQ-MUD, e.g. the bit
error rate, Monte Carlo simulations can be used at the cost of a
large amount of computation, especially when either the number of
users or the search width is large. As an alternative, asymptotic
analysis in the large system limit ($K\rightarrow \infty,
N\rightarrow \infty$ while keeping $\frac{K}{N}=\beta$) can be
applied in view of the recent success of such analysis in
evaluating individual optimal MAP multiuser detection (IO-MUD)
~\cite{Tanaka2002}~\cite{Caire2003}.

For simplicity, we consider only the case of $L=L_1=2$, as
introduced in Section IV.E, and focus on the detection of the
$t$-th symbol of user $k$. Since the power of the desired signal
is distributed in two successive symbol periods, we can consider
symbol $b_k(t)$ as being transmitted through two independent
virtual channels, detected independently and combined to obtain
the soft input to the decoders. In each channel, there exist two
groups of interferers, each of which contains $\kappa K$ users:
one from the MAI in the current symbol period and the other from
the ISI in the previous symbol period. After despreading, the
powers of the desired signal, interference and noise, which are
denoted by $\left\{Q_i\right\}_{i=0,1}$,
$\left\{C(i,j)\right\}_{i,j=0,1}$ and
$\left\{N(i)\right\}_{i=0,1}$, respectively, are illustrated in
Fig.3, where channel 0 and 1 denote the symbol period $t$ and
$t+1$, respectively. We can treat the ISI as MAI from $\kappa K$
virtual users, thereby unifying the ISI into the framework of MUD.

Due to the assumption of exponential decaying variance of channel
gains in (8), we can obtain the parameters in the equivalent
channel with explicit expressions, which are given by
\begin{eqnarray}
  C(i,j) &=&E\left\{|\rho_{kn}^{t\tau}(i,j)|^2\right\}\nonumber\\
         &=&\left\{\begin{array}{lll}
         \frac{e^{-\lambda(i+j)}(1-e^{-2\lambda})(e^{\lambda}-1)^2}{2\lambda},  &\qquad{\mbox{if $i>0$ and $j>0$}} \\
         e^{-\lambda j}(e^{\lambda}-1)\left(\frac{1-e^{-\lambda}}{\lambda}-\frac{1-e^{-2\lambda}}{2\lambda}\right),    &\qquad{\mbox{if $i=0$ and $j>0$}}                                    \\
         \left(1-\frac{2(1-e^{-\lambda})}{\lambda}+\frac{1-e^{-2\lambda}}{2\lambda}\right),  &\qquad    \mbox{if $i=j=0$}
         \end{array}\right.,
\end{eqnarray}
\begin{eqnarray}
Q(i)&=& E\{|\rho^{tt}_{kk}(i,i)|^2\}\nonumber\\
    &=&\left\{\begin{array}{ll}\left(1-\frac{1-e^{-\lambda}}{\lambda}\right),\qquad &\mbox{if  }i=0\\
       \frac{e^{-\lambda(i-1)}(1-e^{-\lambda})^2}{\lambda},\qquad       &\mbox{if  }i>0
       \end{array}\right.,
\end{eqnarray}
and
\begin{eqnarray}
N(i)=\sigma_n^2+(1-\kappa)\beta E\{\Delta
b^2\}\left(C(i,i)+C(0,1)\right),\qquad i=0,1,
\end{eqnarray}
where $\beta=\frac{K}{N}$ is the system load.

It is shown in ~\cite{Tanaka2002} that in the large system limit,
the multiaccess channel $i$, $i=1,2$, is equivalent to a single
user AWGN channel with input signal power $\eta_iQ(i)$ and noise
power $N(i)$, where $\eta_i$ is the (non-asymptotic) multiuser
efficiency of the virtual channel $i$. Using the replica method
developed in the statistical mechanics of spin glasses
~\cite{Nishimori2001}, the multiuser efficiency $\eta_i$ can be
obtained by solving the following equation ~\cite{Caire2003}:
\begin{eqnarray}
\frac{1}{\eta_i}=1+\sum_{j=0}^1\kappa\beta
E_{\hat{b}}\left\{\gamma_{ij}\left(1-\hat{b}^2\right)
\int_{\mathcal{R}}
\frac{1-\tanh\left(z\sqrt{\gamma_{ij}\eta_i}+\gamma_{ij}\eta_i\right)}
{1-\hat{b}^2\tanh^2\left(z\sqrt{\gamma_{ij}\eta_i}+\gamma_{ij}\eta_i\right)}Dz\right\},
\end{eqnarray}
where
$Dz=\frac{1}{\sqrt{2\pi}}\exp\left({-\frac{z^2}{2}}\right)dz$,
$\gamma_{ij}=\frac{C(i,j)}{N(i)}$, and the expectation is taken
over the distribution of soft decision feedback, which can be
obtained by Theorem IV.3.

On obtaining the multiuser efficiencies of both virtual channels,
we can compute the multiuser efficiency after combining the
results from both channels by
\begin{eqnarray}
\eta=\frac{\sum_{i=0}^1\frac{\eta_iQ(i)}{N(i)}}{\frac{1}{\sigma_n^2}}.
\end{eqnarray}

By obtaining the distribution of the soft decoder output by
simulation, the evolution of $\eta$ with respect to the iteration
stages can be achieved. This results in the evolution of bit error
rate as a byproduct.
\subsection{Parallel interference cancellation}
If we set $\kappa=0$, the RS EQ-MUD degenerates to PIC, which
treats all decision feedbacks as reliable and eliminates the time
consuming trellis search at the cost of some performance
degradation. The multiuser efficiency of PIC is given by
\begin{eqnarray}
\eta=\frac{\sum_{i=0}^1\frac{Q(i)}{N(i)}}{\frac{1}{\sigma_n^2}}.
\end{eqnarray}

\section{Numerical Results}
\subsection{Bit error rate}
Figure 4 shows the bit error rate for different values of
$\frac{E_b}{N_0}$, where $E_b$ denotes the energy per information
bit and $N_0=2\sigma_n^2$ is the noise one-sided spectral density,
under the conditions $K=30$, $\kappa=0.1$ and $\beta=1$. The
iteration times are labelled near the corresponding curves. The
channel codes are assumed to all be the convolutional code
$(23,33,37)_8$ with constraint length 5 here, and in subsequent
simulations.

We can see that the bit error rate diverges when $\frac{E_b}{N_0}$
is less than 2.5dB. If $\frac{E_b}{N_0}$ is larger than this
threshold, each iteration improves the performance, which
converges to the single user performance with moderate
$\frac{E_b}{N_0}$.

\subsection{Validity of the asymptotic analysis}
Figure 5 shows the comparison between the asymptotic analysis and
simulation results with different system loads, where
$\frac{E_b}{N_0}=4$dB and the other configurations are the same as
in Fig. 4. We can see that the asymptotic analysis matches the
simulation results of the finite case quite well, even when the
unreliable set is small (here $\kappa K=3$). Thus in the future
experiments, we apply the asymptotic analysis for computational
efficiency.

\subsection{Performance with different search widths and system loads}
Figure 6 shows the bit error rate with the system load $\beta$
ranging from 0.8 to 1.2 and with the search width $\kappa=0$
(namely, PIC), $0.2$ and $1$ (namely, optimal MAP EQ-MUD). The
iteration times required for convergence are labelled on the
figure. We can see that PIC achieves approximately the same
performance of RS EQ-MUD with $\kappa=0.2$, at the cost of more
iterations. Both PIC and RS EQ-MUD with $\kappa=0.2$ diverge when
$\beta>0.95$, whereas optimal MAP EQ-MUD attains the single-user
performance for almost all system loads with slightly increasing
iteration times. This means that the use of small search widths
may incur considerable performance loss when compared with the
optimal MAP EQ-MUD.

Figure 7 shows the bit error rate with various search widths
$\kappa$ in the two cases of $\beta=1.1$, $\frac{E_b}{N_0}=4$dB
and $\beta=1.2$, $\frac{E_b}{N_0}=5$dB. The required iteration
times are labelled on the figure. We can observe the waterfall
phenomenon near $\kappa=0.5$ and the iteration times are reduced
as the search width increases, at the cost of rapidly ascending
computation. The evolution of multiuser efficiency $\eta$ with
$\beta=1.1$ and $\frac{E_b}{N_0}=4dB$, is given in Fig. 8, where
$\eta$ increases to more than 0.9 with different required
iteration stages when $\kappa>0.5$.

\section{Conclusions}
Mitigation of ISI and MAI is of considerable importance to the
performance of DS-CDMA systems. We have formulated the optimal MAP
EQ-MUD for such systems with channel codes. To reduce the
prohibitive computational cost of MAP EQ-MUD, we have proposed the
RS EQ-MUD by decomposing the trellis and classifying the decision
feedback into reliable and unreliable sets with confidence
metrics. Asymptotic analysis has been used to evaluate the system
performance and is shown to match simulation results for finite
numbers of users. Numerical results show that the RS EQ-MUD can
achieve close to single-user performance with moderate
$\frac{E_b}{N_0}$. With a reasonable search width, RS EQ-MUD
outperforms the conventional PIC with higher user capacity and
fewer required iterations.

\begin{figure}
  \centering
  \includegraphics[scale=1]{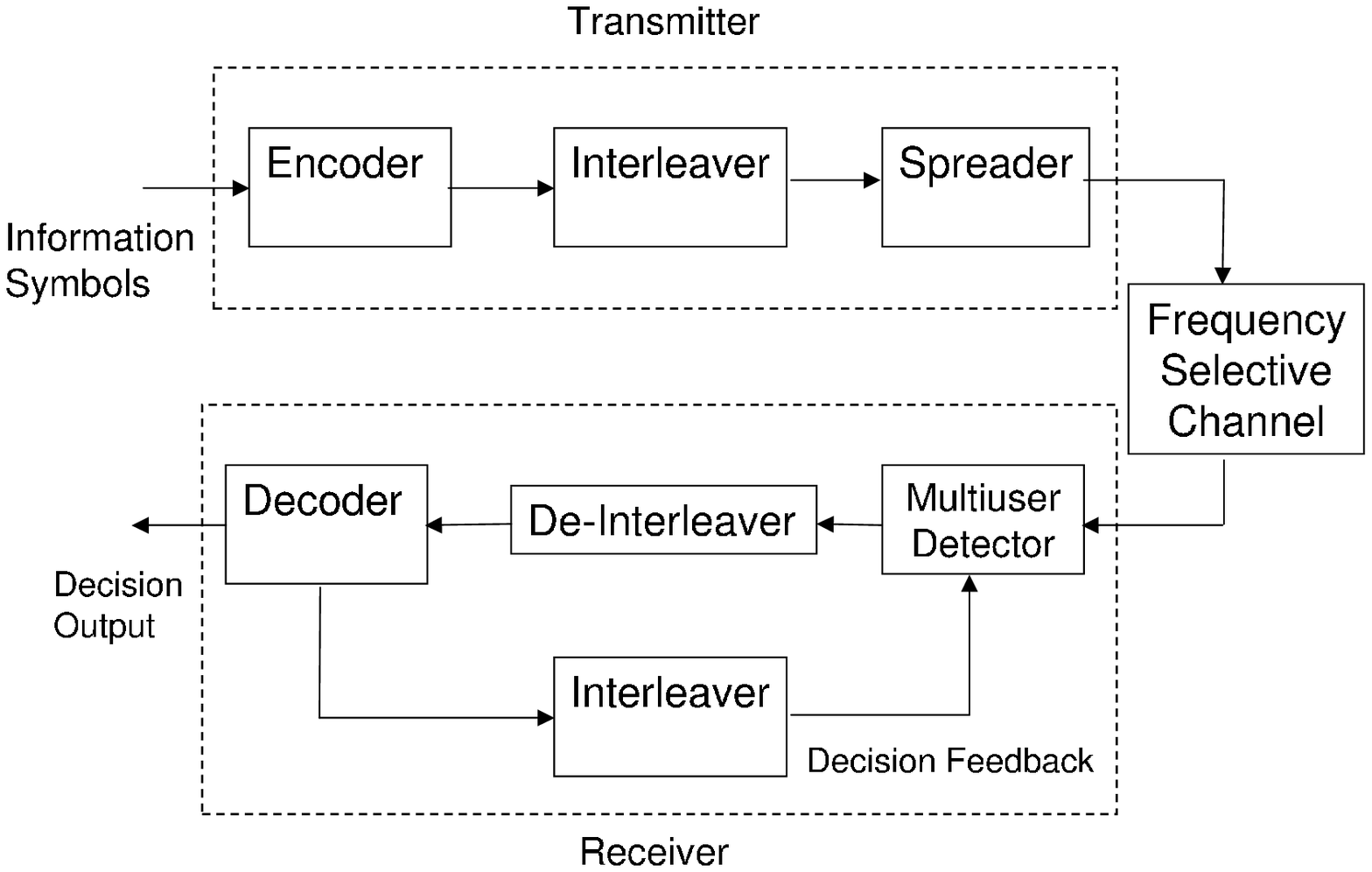}
  \caption{The basic system configuration}\label{}
\end{figure}

\begin{figure}
  \centering
  \includegraphics[scale=1]{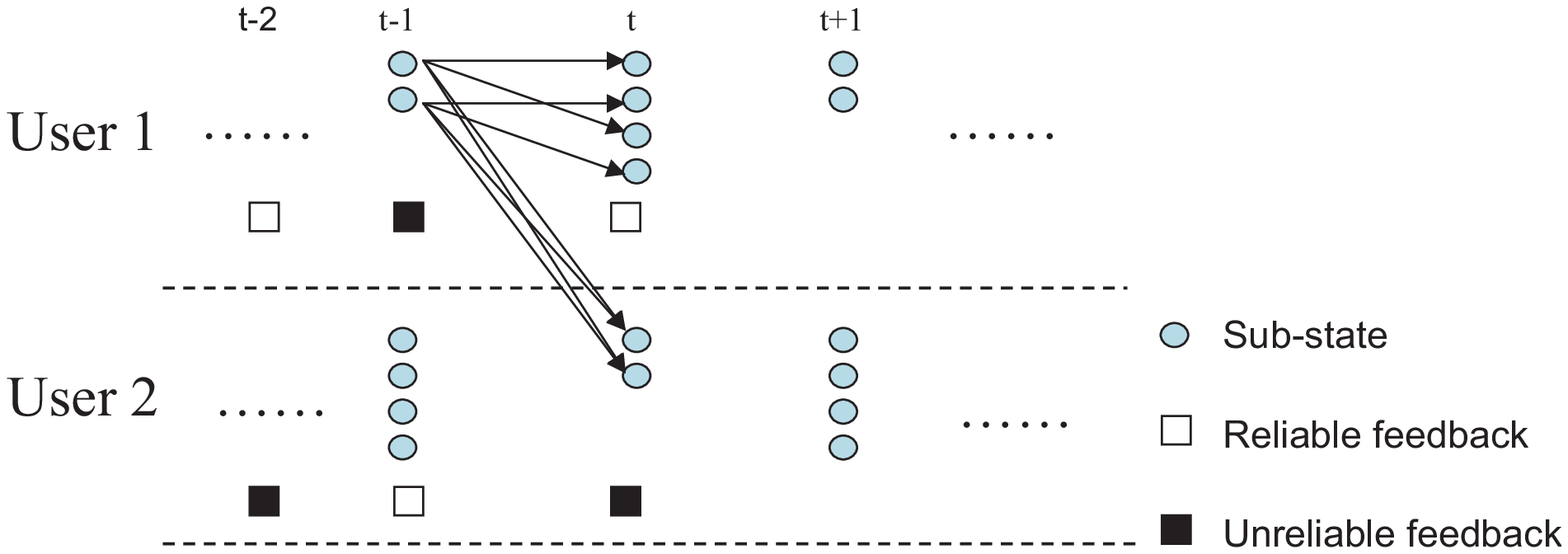}
  \caption{Illustration of two users' sub-trellises}\label{}
\end{figure}

\begin{figure}
  \centering
  \includegraphics[scale=1]{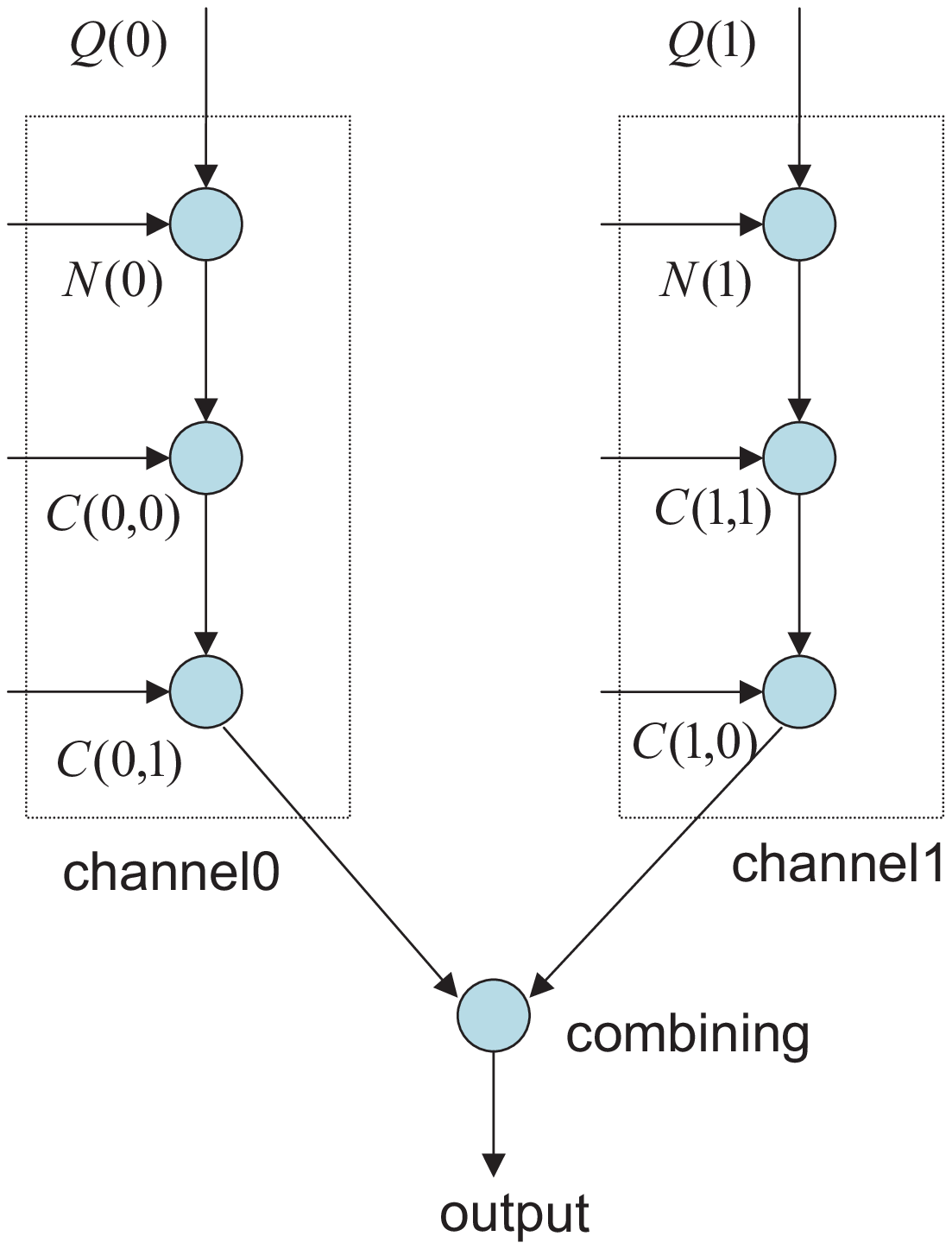}
  \caption{Illustration of the two equivalent independent channels for performance analysis}\label{}
\end{figure}

\begin{figure}
  \centering
  \includegraphics[scale=1]{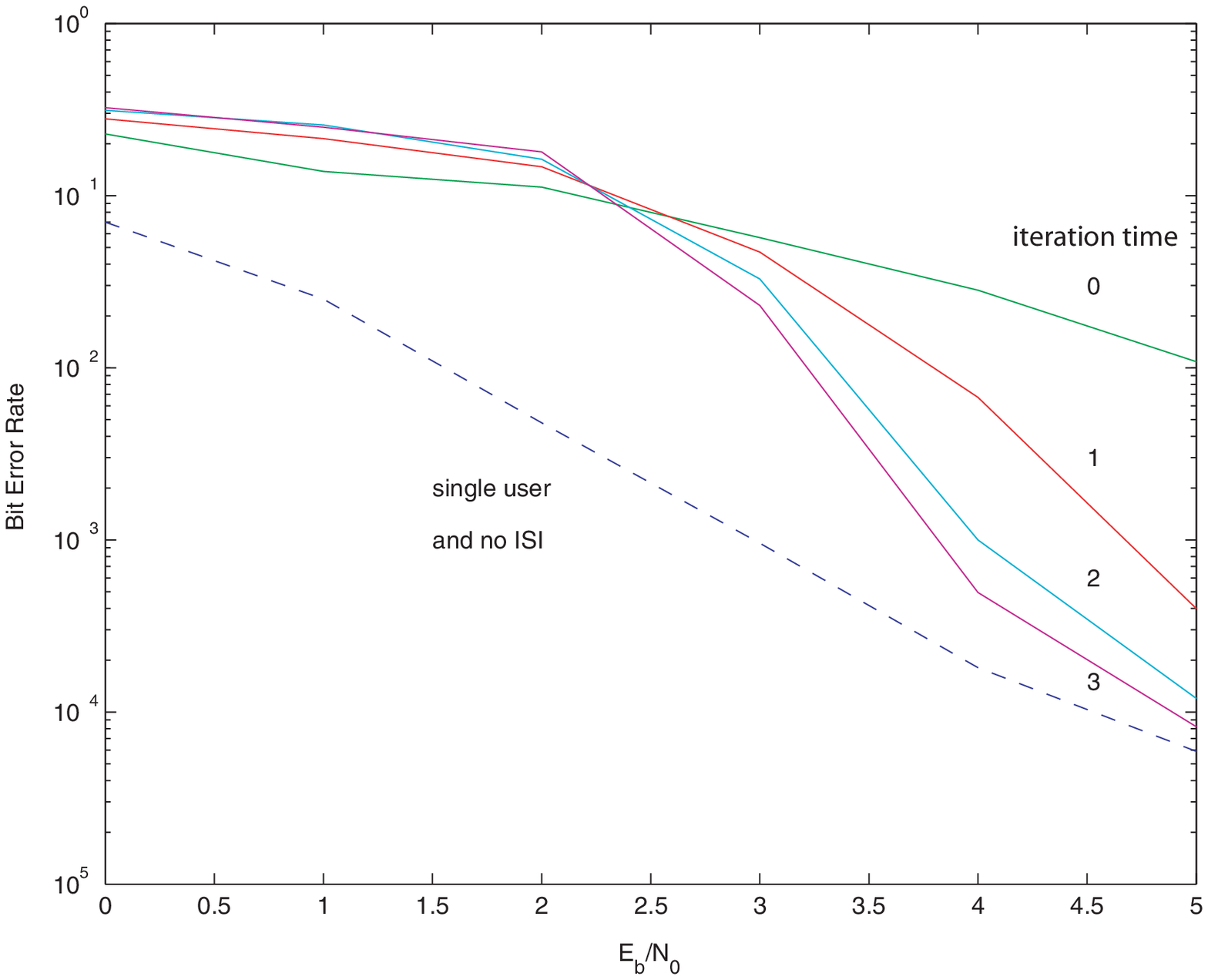}
  \caption{Bit error rate versus $E_b/N_0$}\label{}
\end{figure}

\begin{figure}
  \centering
  \includegraphics[scale=1]{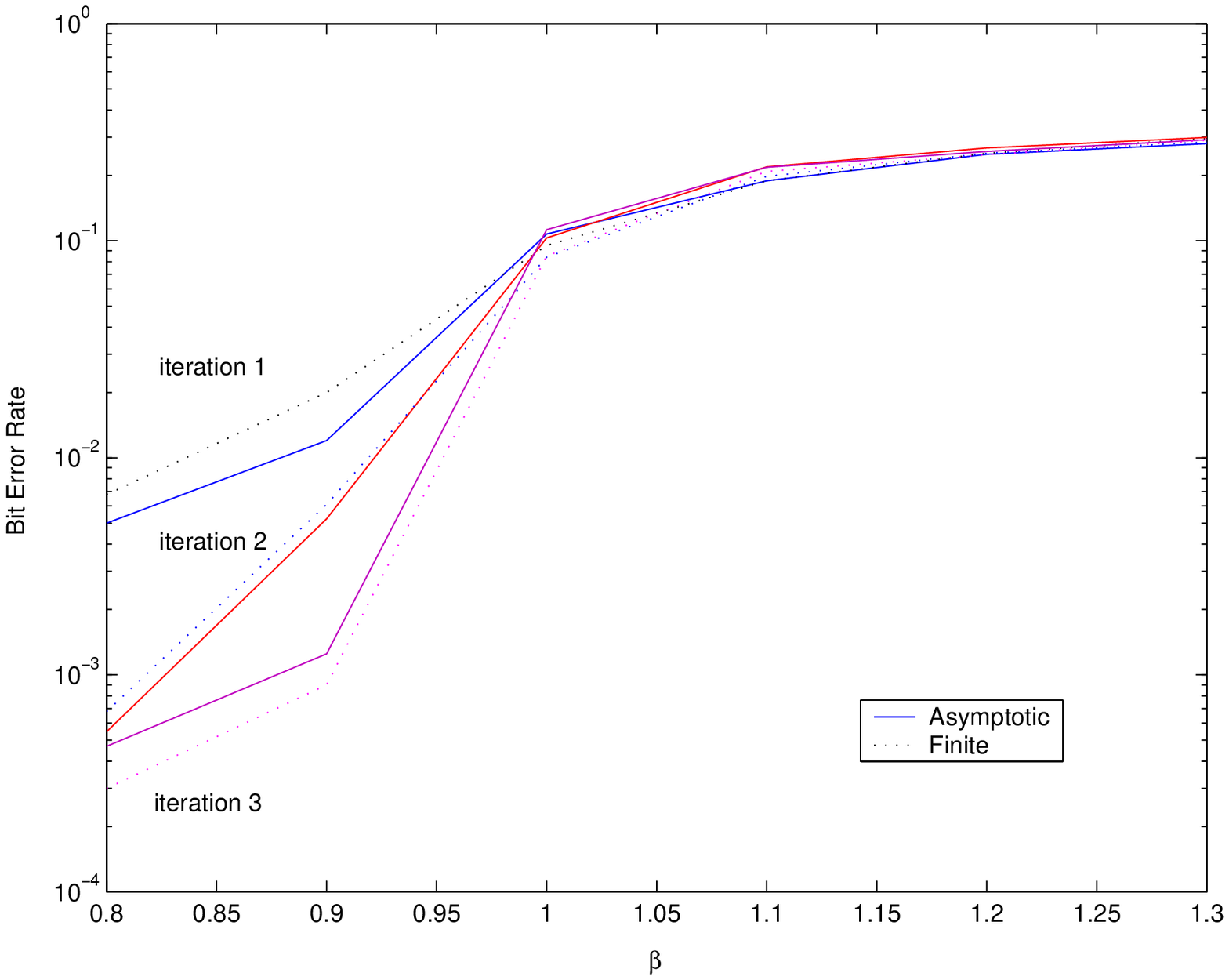}
  \caption{Comparison between asymptotic analysis and simulation results for finite systems}\label{}
\end{figure}

\begin{figure}
  \centering
  \includegraphics[scale=1]{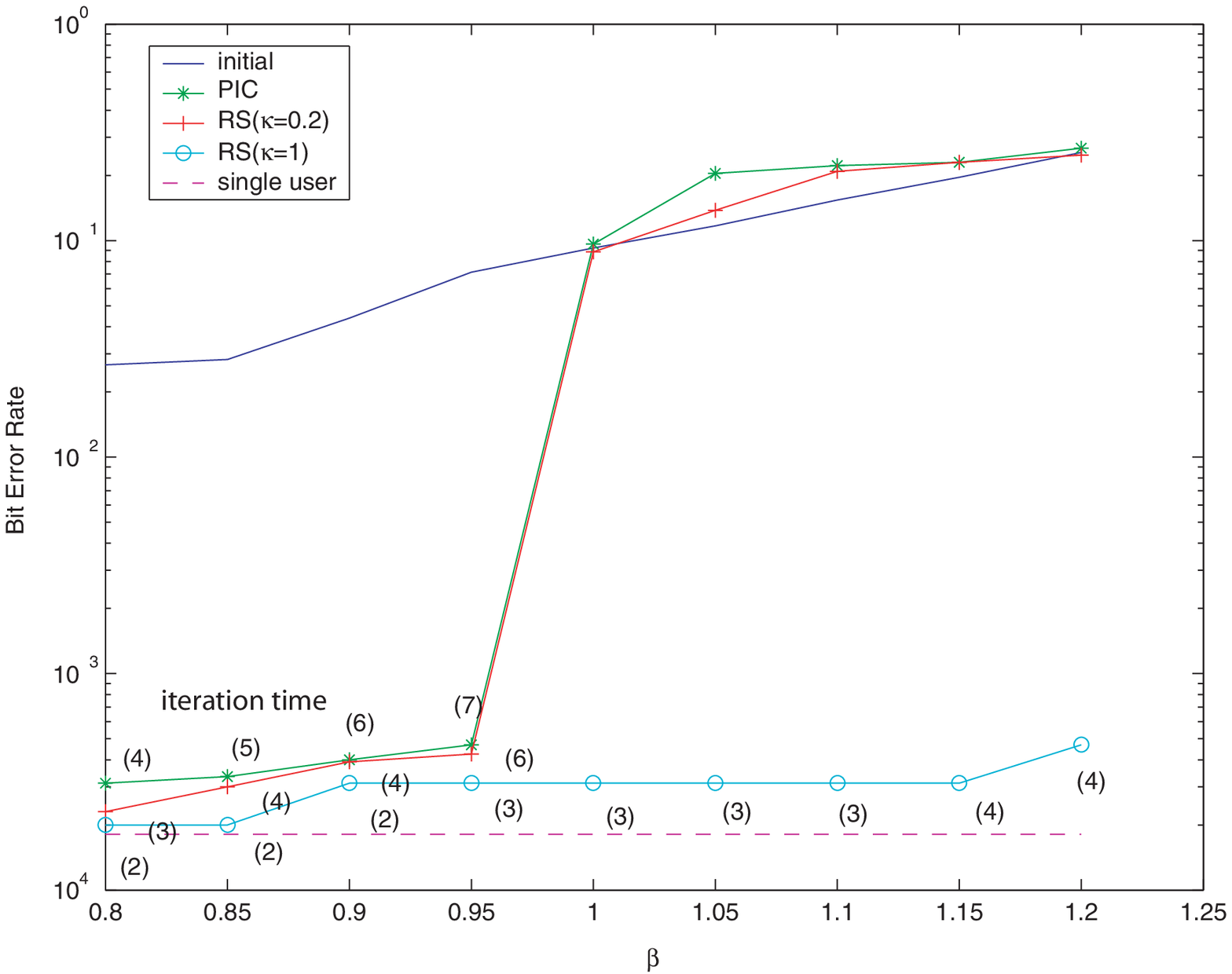}
  \caption{Bit error rate versus system load}\label{}
\end{figure}

\begin{figure}
  \centering
  \includegraphics[scale=1]{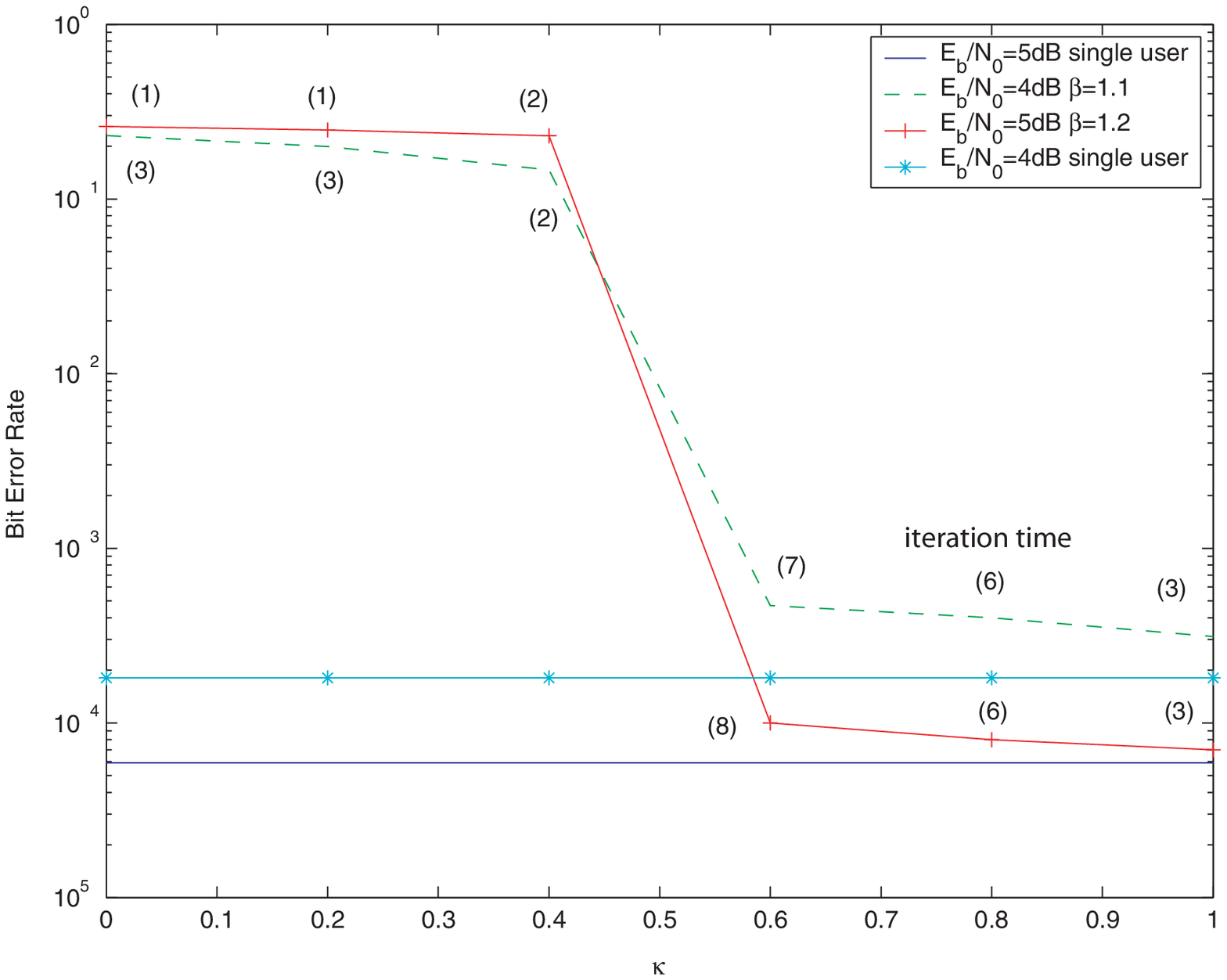}
  \caption{Bit error rate versus search width}\label{}
\end{figure}

\begin{figure}
  \centering
  \includegraphics[scale=1]{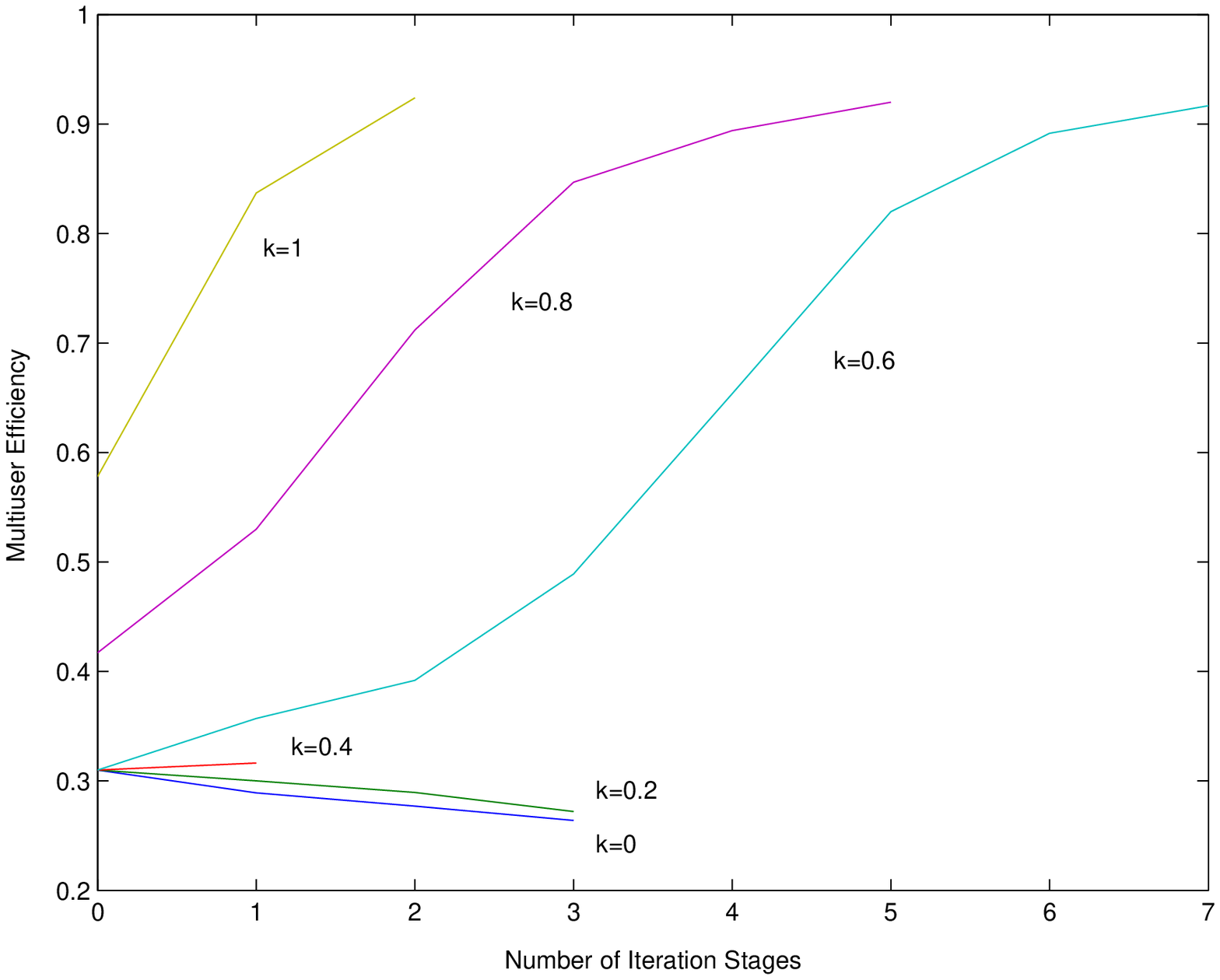}
  \caption{Multiuser efficiency at different iteration stages}\label{}
\end{figure}

\begin{thebibliography}{11}
\bibitem{Alex2000}
P. ~Alexander, A. ~Grant and M. C. Reed, ``Iterative detection on
code-division multiple-acess with error control coding,'' {\em
European Trans. Telecommun.}, Vol.~9,
  pp.~419-426, Aug. 1998.

\bibitem{BCJR}
L. ~R. ~Bahl, J. ~Cocke, F. ~Jelinek and J. ~Raviv, ``Optimal
decoding of linear codes for minimizing symbol error rate,'' {\em
IEEE Trans. Inform. Theory}, Vol.~20,
  pp.~284-287, Aug. 1974.

\bibitem{Beheshti1998}
S. Beheshti, S. H. Isabelle and G. W. Wornell, ``Joint intersymbol
and multiple-access interference suppression algorithms for CDMA
systems,'' {\em European Trans. Telecommunications}, Vol.~9,
  pp.~403-418, Sept./Oct. 1998.

\bibitem{Bokolamulla2003}
D. Bokolamulla and T. Aulin, ``Reduced complexity iterative
decoding for concatenated coding schemes,'' {\em Proceedings of
the 2003 IEEE International Conference on Communications},
Anchorage, Alaska, May 11-15, 2003.

\bibitem{Caire2003}
G. Caire, R. M\"{u}ller and T. Tanaka, ``Iterative multiuser joint
decoding: Optimal power allocation and low-complexity
implementation,'' submitted to {\em IEEE Trans. Inform. theory},
2003.

\bibitem{DuelHallen1989}
A.~Duel-Hallen and C.~Heegard, ``Delayed decision-feedback
sequence estimation,'' {\em IEEE Trans. Commun.}, Vol.~37,
  pp.~428-434, Aug. 1989.

\bibitem{Eyuboglu1988}
M. V. Eyuboglu and S. U. H. Qureshi, ``Reduced-state sequence
estimation with set partitioning and decision feedback,'' {\em
IEEE Trans. Commun.}, Vol.~36,
  pp.~13-19, Jan. 1988.

\bibitem{Fossorier}
M. P. C. Fossorier, ``Iterative reliability-based decoding of
low-density parity check codes,'' {\em IEEE J. Select. Areas
Commun.}, Vol.~19,
  pp.~908-917, May, 2001.

\bibitem{Frey1998}
B. J. Frey and F. R. Kschischang, ``Early detection and trellis
splicing: reduced-complexity iterative decoding,'' {\em IEEE J.
Select. Areas Commun.}, Vol.~16,
  pp.~153-159, Feb. 1998.

\bibitem{Hsu2001}
J. Hsu and C. Wang, ``A low-complexity iterative multiuser
receiver for turbo-coded DS-CDMA systems,'' {\em IEEE J. Select.
Areas Commun.}, Vol.~19,
  pp.~1775-1783, Sept. 2001.

\bibitem{Laot2001}
C. Laot, A. Glavieux and J. Labat, ``Turbo equalization : Adaptive
equalization and channel decoding jointly optimized,'' {\em IEEE
J. Select. Areas Commun.}, Vol.~19,
  pp.~1744-1752, Aug. 2001.

\bibitem{Lupas1989}
R.~Lupas and S.~Verd\'{u}, ``Linear multiuser detectors for
synthronous code-division multiple-acess channels,'' {\em IEEE
Trans. Inform. Theory}, Vol.~35,
  pp.~123--136, Aug. 1989.

\bibitem{Muller1996}
S. H. M\"{u}ller, W. H. Gerstacker and J. B. Huber,``Reduced-state
soft-output trellis-equalization incorporating soft feedback,''
{\em Proceedings of IEEE Global Telecommunications Conference },
London, UK, Nov. 1996.


\bibitem{Nishimori2001}
H. Nishimori, {\em Statistical Physics of Spin Glasses and
Information Processing}.
\newblock Oxford University Press, UK, 2001.

\bibitem{Proakis2001}
J. G. Proakis, {\em Digital Communications (fourth edition)}.
\newblock McGraw Hill, USA, 2001.

\bibitem{Qin2002}
Z. Qin, K. C. Teh and E. Gunawan, ``Iterative reduced-state
multiuser detection for asynchronous coded CDMA,'' {\em IEEE
Trans. Commun.}, Vol.~50,
  pp.~1892-1894, Dec. 2002.

\bibitem{Raphaeli2000}
D. Raphaeli and T. Kaitz, ``A reduced-complexity algorithm for
combined equalization and decoding,'' {\em IEEE Trans. Commun.},
Vol.~48,
  pp.~1797-1807, Nov. 2000.

\bibitem{Rappaport1999}
T. S. Rappaport, {\em Wireless Communications : Principles and
Practice}.
\newblock Prentice Hall, Upper Saddle River, NJ, USA, 1999.

\bibitem{Reed1997}
M. C. Reed, C. Schlegel, P. D. Alexander, and J. A. Asenstorfer,
``Reduced complexity iterative multiuser detection for CDMA with
FEC,'' {\em Proceedings IEEE International Conference on Universal
Personal Communications}, San Diego, USA, Oct 12-16th 1997.

\bibitem{Shamai2002}
S. ~Shamai (Shitz) and S. ~Verd\'{u}, {"Decoding Only the
Strongest CDMA Users",}. In \newblock {\em Codes, Graphs, and
Systems}, R. Blahut and R. Koetter, Eds., pp. 217-228, Kluwer,
2002.

\bibitem{Tanaka2002}
T. Tanaka, ``A statistical-mechanics approach to large-system
analysis of CDMA multiuser detectors,'' {\em IEEE Trans. Inform.
theory}, Vol.~48,
  pp.~2888--2910, Nov. 2002.

\bibitem{Tuchler2000}
M. T\"uchler, R. Kotter and A.~Singer, ``"Turbo equalization :
Principles and new results,'' {\em IEEE Trans. Commun.}, Vol.~50,
  pp.~754-767, Aug. 2002.

\bibitem{wang1999}
X. Wang and H. V. Poor, ``Iterative(Turbo) soft interference
cancellation and decoding for coded CDMA,'' {\em IEEE Trans.
Commun.}, Vol.~47,
  pp.~1046-1061, Aug. 1999.

\bibitem{Verdu1998}
S. Verd\'{u}, {\em Multiuser Detection}.
\newblock Cambridge University Press, Cambridge, UK, 1998.

\bibitem{Young1996}
S. Young, ``A review of large vocabulary continuous-speech
recognition,'' {\em IEEE Signal Processing Magazine}, Vol.~10,
  pp.~133-145, March. 1996.

\end{thebibliography}
\end{document}

% END of IEEEtest.tex ************